An *ab initio* study of intrinsic defects in zirconolite

Jack Mulroue, Andrew J. Morris and Dorothy M. Duffy

Department of Physics and Astronomy and the London Centre for Nanotechnology, University College London, Gower Street, London, WC1E 6BT

**Abstract**
Zirconolite, $CaZrTi_2O_7$, is a proposed ceramic for the use in disposal of highly active nuclear waste. Density functional theory (DFT) has been used, in conjunction with a random search technique, to identify the stable interstitial sites for the intrinsic defects in zirconolite. There is a significant dependence on charge states for the vacancy defect structures, with the formation of an $O_2$ molecule in certain charge states of Ti and Zr vacancies. The low coordination of the oxygen ions provided sites on which DFT localised holes associated with the cation vacancies. The Frenkel defect energies, calculated from the combination of the vacancy and interstitial formation energies, show that the oxygen defects have a strong dependence on the chemical environment, with a range of 1.5 eV between the oxygen Frenkel pair energies.

**1. Introduction**
The proposed method for the long-term storage of high-level radioactive waste is the encapsulation of the waste in a ceramic matrix, followed by storage in an underground repository.[1] One such ceramic, which has many of the properties thought to be favourable for waste storage, is zirconolite ($CaZrTi_2O_7$). Zirconolite is one of the components of SYNROC, a synthetic rock composed of a number of ceramics which is able to incorporate the long lived radioactive actinide species (Pu,U).[2-4]

The decay of the encapsulated radioisotopes in a ceramic will modify the microstructure by the creation of defects and, possibly, amorphisation. As well as the lattice displacements, the alpha particle emitted from these elements during decay will results in a large number of excited electrons. The alpha particle dissipates its energy in to the electronic system producing excited electrons and hole, which are able to trap at defect sites in ceramic materials. It has been shown recently that trapping of holes and electrons at defects in MgO has a significant effect on their properties.[5] The long timescale durability of the material will depend on the evolution of the microstructure over periods of hundreds or even thousands of years, well beyond the timescale of experimental measurements. A detailed understanding of the properties of defects is, therefore, a necessary starting point for the prediction of the long-term durability of the encapsulating ceramics.

There have been few modelling studies of zirconolite to date due to the complex crystal structure. Classical molecular dynamics simulations of low energy cascades have been performed on uranium loaded zirconolite and displacement energies have been calculated.[6] Experimental analysis of the material has included electron microscopy of natural zirconolite containing actinide species,[7] X-ray diffraction (XRD) of synthesised zirconolite containing Pu or its surrogate Ce,[8-10] ion beam studies,[2] X-ray absorption studies (XAS),[11] leaching studies[12,13] and the behaviour of He inside the crystal.[14,15]

In this paper we use density functional theory (DFT) to calculate the fundamental properties of intrinsic point defects (vacancies and interstitials) in zirconolite. The complex crystal structure of zirconolite makes it challenging to predict the stable configurations of interstitial ions, therefore we use the *ab initio* random structure search (AIRSS) technique. AIRSS has been used successfully for locating novel crystal structures[16-18] and has been shown to be proficient in deducing point defect structures.[19,20] We investigate the effects of trapped charge on the defect conformations and formation energies.

## 2. Computational Method

Zirconolite has a monoclinic crystal structure consisting of rows of 6 and 4 fold coordinated Ti-O polyhedra separated by alternating layers of Ca and Zr ions (figure 1).

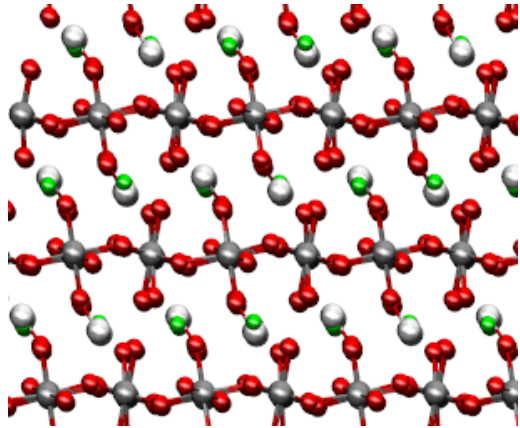

*Figure 1: (Color online) Structure of bulk zirconolite projected in the [010] direction. Red (small dark) spheres show oxygen ions, gray (large gray) spheres show Ti ions, white spheres show Zr ions and green (small gray) spheres show Ca ions.*

The crystal structure was obtained from X-ray diffraction experiments performed by Rossell.[21] Zirconolite can exist in a number of different stoichiometries,[21,22] $CaZr_xTi_{3-x}O_7$, where $0.80 < x < 1.37$, but a value of x=1 has been used in this research to minimize the size of the simulation cell. Experimental studies of zirconolite have detected a 5 fold coordinated Ti ion instead of the 4 fold coordinated Ti ion in the ideal structure. Our modelling studies suggest that this could be related to the non-stoichiometry of the experimental sample. Substituting one Zr ion per unit cell for Ti (x = 0.875) results in the substitutional Ti ion becoming 5 fold coordinated. A unit cell with an excess of Zr (x = 1.125), created by substituting one Ti ion per unit cell for a Zr ion, results in the two 4 fold coordinated Ti polyhedra in the same layer as the extra Zr ion becoming 5 fold coordinated. Therefore $CaZr_{1.25}Ti_{1.75}O_7$ must have the excess two Zr ions in different Ti layers to remove the 4 fold coordinated Ti polyhedra.

The periodic plane wave DFT calculations were performed using the VASP code[23,24] with Perdew-Burke-Ernzerhof (PBE) exchange-correlation functional,[25] Projector Augmented Wave (PAW) pseudopotentials and a single *k* point sampling of the Brillouin zone at the $\Gamma$ point, corresponding to a minimum Brillouin zone sampling of $0.07 Å^{-1}$ with a basis set containing plane waves up to an energy of 600 eV. A $1 \times 2 \times 1$ supercell containing 176 ions was used in this work. Relaxation of the cell volume resulted in relaxed lattice parameters that are in reasonable agreement with those obtained by Rossell using XRD experiments, (table I).[21] The defects were then introduced into this cell and the structure relaxed with the lattice parameters fixed.

*Table I: Lattice parameters calculated by DFT and compared to experimental XRD values obtained by Rossell.[21]*

| Lattice Parameter | DFT PBE (Å) | Experimental (Å) | % error |
|---|---|---|---|
| a | 12.09 | 12.45 | 2.89 |
| b | 14.14 | 14.55 | 2.82 |
| c | 11.08 | 11.39 | 2.72 |
| Volume (Å$^3$) | 1862.82 | 2028.12 | 8.15 |

## 2.1 Interstitial defects

The complicated crystal structure of zirconolite presents problems for modelling interstitial point defects because it is difficult to predict the stable configurations. In this study an efficient random search technique (AIRSS) was employed to identify energy minima. Interstitial atoms were placed randomly within the simulation cell and the ionic positions are structurally relaxed within VASP. Once a local minimum is located the process is repeated until the required number of minima are found, then the local minima are sorted by energy. The lowest energy interstitial sites were identified from 100 random searches for each interstitial defect. This involved insertion of an additional atom, with the number of electrons corresponding to the desired charge state, at randomly generated coordinates, in an 88 ion unit cell. The 89 ions were relaxed until all the forces were less than to 0.1 eVÅ$^{-1}$ using the pseudopotentials Ca(3p$^6$ 4s$^2$), Zr(4d 5s 5p), Ti(3d$^3$ 4s$^1$) and O(2s$^2$ 2p$^4$, with a maximum of 283 eV of kinetic energy used in the pseudopotential) to describe the atoms. The 20 lowest energy configurations resulting from the 100 searches were identified and these were further relaxed until the forces were less than 0.01 eVÅ$^{-1}$. The lowest energy interstitial configuration was identified from these 20 searches. The interstitial was inserted into the larger 176 ion supercell and the ions relaxed until the forces were less than 0.01 eVÅ$^{-1}$ using the pseudopotentials Ca(3s$^2$ 3p$^6$ 4s$^2$), Zr(4d 5s 5p), Ti(3d$^3$ 4s$^1$) and O(2s$^2$ 2p$^4$, with a maximum of 700 eV of kinetic energy used in the pseudopotentials) to describe each ion. The interstitial energies from this larger cell were used to calculate the Frenkel defect energies.

## 2.2 Vacancies

The vacancies were created by removing the corresponding lattice ion from the 176 ion cell before relaxing all the ions, with the lattice vectors fixed, until the forces were less than 0.01 eVÅ$^{-1}$ on each atom. The number of electrons in the cell was adjusted to obtain the required oxidation state. Each vacancy species was studied with a charge ranging from 0 to the formal charge of the vacancy (i.e. for the Ti vacancy the Ti atom was removed and 0, 1, 2, 3 or 4 electrons added to the cell). The VASP code introduced a neutralising background charge when the system was not charge neutral.

## 3. Results and discussion

The band gap calculated for zirconolite in the current simulations was 2.8 eV, which is 0.8 eV less than the experimentally obtained value of 3.6 eV.[26] This is in line with the general tendency of DFT to underestimate band gaps in insulators.[27,28] A study of the partial density of states has shown that the edge of the valence band is constructed from the p orbitals from the oxygen ions, while the conduction band edge is constructed from Ti d orbitals (figure 2).

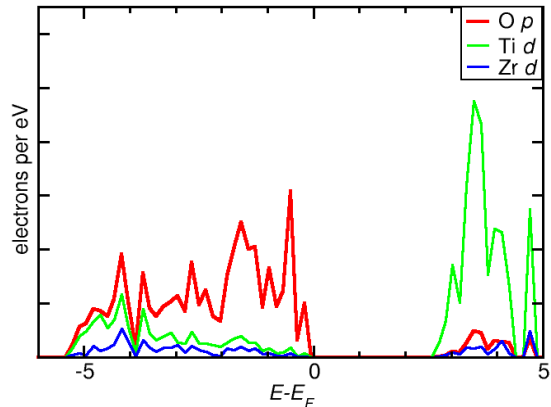

*Figure 2: (Color online) Partial density of states of bulk zirconolite. Red (thick dark) show oxygen p orbitals, green (light thin) show titanium d orbitals and blue (dark thin) show zirconium d orbitals.*

### 3.1 Interstitials

The lowest energy configuration of the neutral oxygen interstitial is dependent on the size of the simulation cell and, therefore, the defect concentration. A <111> dumb-bell interstitial was the lowest energy configuration identified within the 88 ion cell, with a lattice oxygen from a six fold coordinated Ti polyhedron forming the second atom of the split interstitial (figure 3a). The oxygen dumb-bell has also been identified as the lowest energy configuration for a neutral oxygen interstitial in a number of oxides.[5,29-31] In zirconolite the oxygen atoms of the dumb-bell have a separation distance of 1.5 Å and Bader analysis[32] of the relaxed structure shows that -0.48e is transferred from the lattice oxygen to the interstitial, resulting in a charge of -0.57e on both oxygen atoms of the dumb-bell. The formation of the oxygen dumb-bell creates a defect level which resides 0.29 eV above the valence band maximum (figure 3c). However, a different interstitial configuration was formed in the 176 ion cell. In the larger cell the interstitial results in the increase in the coordination of two 6 coordinated Ti ions to 7 fold coordinated, with the maximum Ti-O bond length defined as 2.2 Å. The interstitial also increased the coordination of all the 4 fold coordinated Ti, with three increasing to 6 fold coordination and the remaining five to 5 fold coordination (figure 3b). Two unoccupied defect levels form in the band gap, 0.40 eV and 0.69 eV above the valence band for this defect configuration (figure 3d).

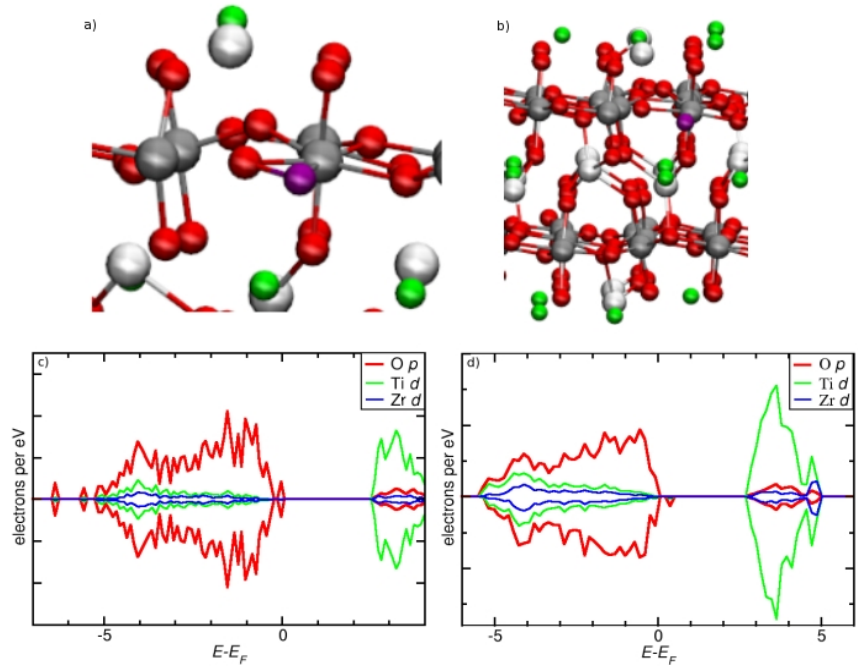

*Figure 3: (Color online) The relaxed configuration of the $O^0$ interstitial in a) a 88 unit cell and b) a 176 supercell. The purple (large dark) spheres show the interstitials ion. Partial density of states from a) the 88 unit cell and d) the 176 supercell. The upper and lower parts for this, and subsequent, density of states plot represent values for spin up and spin down electrons respectively.*

AIRSS located a defect structure for the singly charged oxygen interstitial that was not dependent on the cell size. The oxygen ion binds to a 4 fold coordinated Ti polyhedron producing a defective 5 fold coordinated Ti polyhedron (figure 4a). Bader analysis shows that no significant electron density is redistributed due to these displacements. The defect level created by the $O^-$ interstitial is 0.54 eV above the valence band (figure 4b). The doubly charged oxygen interstitial was found to be unstable, as it relaxed to the same configuration as the singly charged interstitial, with the extra electron delocalised over the cell.

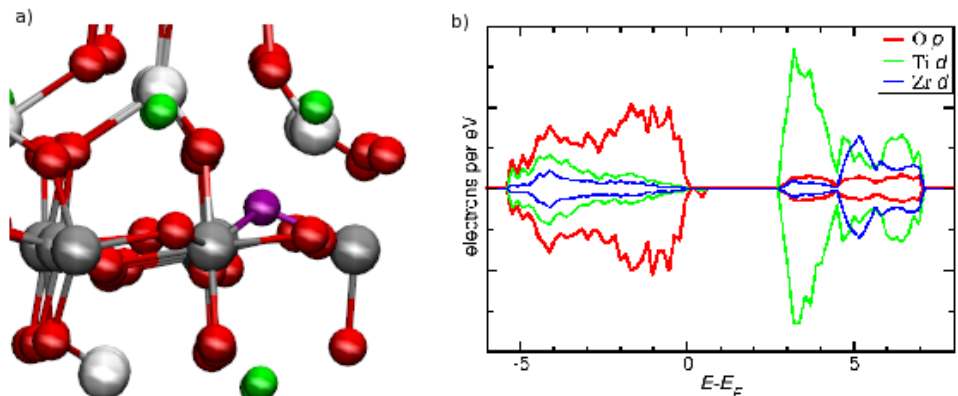

*Figure 4: (Color online) a) The relaxed configuration of the $O^-$ interstitial in the 176 ion supercell and b) the partial density of states.*

The calcium interstitial was inserted in two different charge states, the neutral and +2e, but Bader analysis of the resulting configuration shows that the neutral interstitial lost its electrons, giving a charge state of +1.4e, indicating that the calcium interstitial is only stable in the doubly charged state. The double charged interstitial also had a charge of +1.4e, 0.2e less than the lattice Ca ions. The Ca interstitial resides in the <010> channels which run through zirconolite (figure 5a). The formation of the calcium interstitial does not result in the formation of a defect level in the band gap of zirconolite, because the valence and conduction bands are not constructed from Ca orbitals.

The titanium interstitial was also studied in the neutral and the formal charge (+4e) states, however in both of these cases the electron distribution was such that the resulting interstitial had a charge of +2e, with the neutral interstitial electron density equal +1.4e and the quadruply charged +1.8e. AIRSS was then performed on the doubly charged Ti interstitial and the stable interstitial site was found to be in the <010> channels (figure 5b) close to the position of the Ca interstitial, with a Bader charge of +1.6e. The Ti interstitial results in the formation of two singly occupied defect levels in the band gap, 2.04 eV and 2.23 eV above the valence band (figure 6).

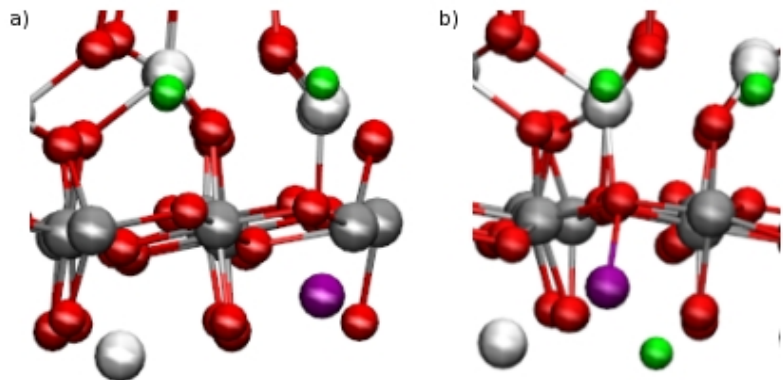

Figure 5: (Color online) The relaxed configuration of a) the $Ca^{2+}$ and b) the $Ti^{2+}$ interstitials located in the <010> channel.

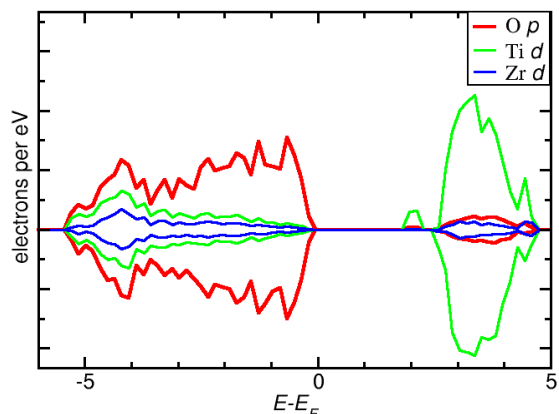

Figure 6: (Color online) Partial density of states of the $Ti^{2+}$ interstitial.

A similar procedure for the zirconium interstitial found that it too exists only in the doubly charged oxidation state, with Bader charge of +1.7e and 2.2e for the neutral and quadruple charged interstitials respectively. However, in this case AIRSS located an interesting lowest energy structure. The additional zirconium ion displaces a 6 fold coordinated Ti ion from its lattice site and resides on the Ti site. The displaced Ti lattice ion becomes an interstitial ion (figure 7). This displaced Ti ion resides in the same position as the Ti interstitial formed above and both have the same Bader charge of +1.61e. The bond distances within the polyhedron increase with the substitution of a Zr for a Ti ion, with one of the O ions displaced away from the polyhedron causing the Zr to be 5 fold coordinated. The substitution of a Ti ion by a Zr ion has been observed previously in pyrochlores[33] and Zr doped $TiO_2$.[34] The substitutional Zr/ Ti interstitial configuration is 0.27 eV more stable than the next lowest energy configuration, which is the Zr interstitial located in the <010> channel.

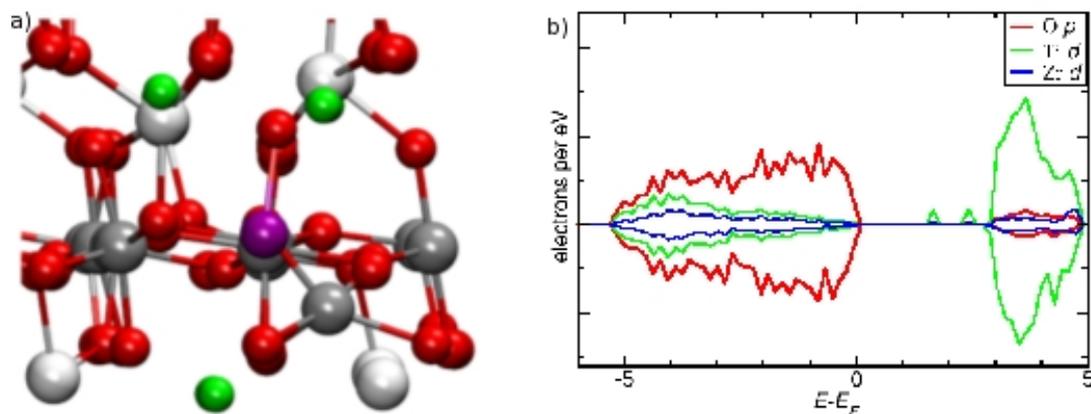

Figure 7: (Color online) (a) The relaxed configuration of the $Zr^{2+}$ interstitial, where the Zr ion has displaced a lattice Ti ion into the <010> channel, (b) the partial DOS for the defect configuration.

### 3.2 Vacancies

There are seven different oxygen chemical environments in zirconolite and vacancies in four of these environments have been studied here, to observe trends in the oxygen vacancy behaviour. The four environments studied were an oxygen ion which bridges two 6 fold coordinated Ti polyhedra in the <010> direction, an oxygen ion which bridges a 6 fold coordinated Ti polyhedron and a Zr ion (1.99 Å Zr-O), another oxygen ion from a 6 fold coordinated polyhedron bridging a Zr ion (2.14 Å Zr-O) and finally a 3 fold coordinated oxygen ion which bridges two 6 fold coordinated polyhedra in the <100> direction and a Zr ion. The local environment of the vacancy had little effect on the localisation of the excess electrons generated from the neutral vacancies, as one of the electrons localised on the Ti ion that the oxygen was bonded to and the other localised on a Ti ion in the same or in an adjacent row. This localisation of the excess electrons is similar to that found in anatase ($TiO_2$).[35] The exception to this was the fourth environment in which one electron localised between the neighbouring Ti ion in an adjacent row and a Zr ion and the other electron localised on another neighbouring Ti (figure 8). The localisation of the excess electron from the singly charged vacancy is consistent for the different environments, where the electron localises on the Ti ion that the oxygen was coordinated to. The second environment is an exception to this behaviour as the excess electron localises between the Ti and Zr ions where the oxygen ion was removed from. Analysis of the electronic density of states (DOS), calculated using the Gaussian smearing approximation, for the oxygen vacancies found that only the neutral defects result in the creation of a defect level in the band gap. The defect levels were constructed from Ti d orbitals with a contribution from the Zr d orbitals for the fourth environment. The energy of these defect levels are 2.66 eV, 2.25 eV and 2.01 eV above the valence band for the first, second and fourth oxygen (figure 9) environment respectively. A single defect level is also found for a neutral oxygen vacancy in anatase.[35]

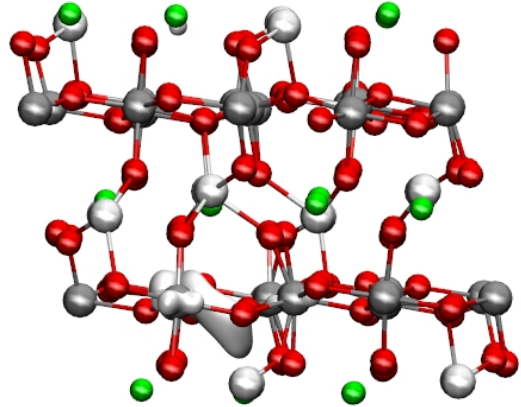

*Figure 8: (Color online) The 0.06 e Å⁻³ isosurface of the $V_O^0$ in the fourth environment.*

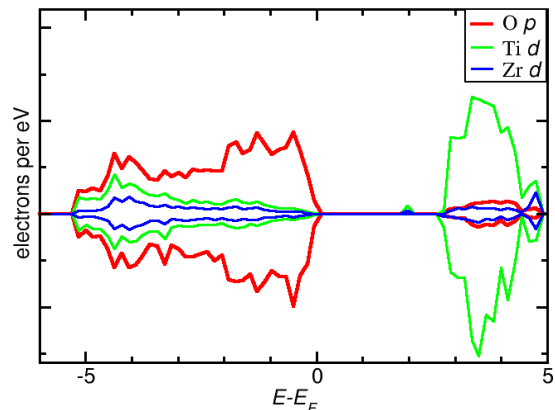

*Figure 9: (Color online) Partial density of states of the neutral oxygen vacancy in the fourth environment.*

The neutral calcium vacancy localised the two holes present on two oxygen ions. One of the oxygen ions is next to the vacancy site, while the other oxygen ion is the top of a 6 fold coordinated polyhedron, in which the bottom of the polyhedron is next to the vacancy site. Both of these oxygen ions are 2 fold coordinated. There are no defect levels created in the band gap due to the presence of the vacancy. The singly charged vacancy also localises the hole state on a 2 fold coordinated oxygen. The localisation of the hole does not result in a defect level being created in the band gap nor is a defect level present for the doubly charged vacancy.

The zirconium vacancy showed significant charge state dependence on the observed defect structure, summarised in table II. The neutral defect results in the formation of an oxygen molecule below the vacancy site. The bond distance of this $O_2$ molecule was 1.24 Å, which compares to 1.22 Å for a free $O_2$. The phonon frequency of this molecule is 37.99 THz compared to the 45.18 THz for the free molecule (1267.10 cm⁻¹ and 1507.09 cm⁻¹ respectively) showing that the molecule has a weaker covalent bond than the free $O_2$ molecule. Analysis of the isosurface showed that there was electron density in between the two oxygen atoms, similar to that shown in figure 11, indicating that a covalent bond has formed. The vacancy also leads to the displacement by 0.83 Å of a 4 fold coordinated Ti polyhedron out of the Ti row towards the vacancy. The defect structure results in the creation of a defect level 2.28 eV above the valence band constructed from Ti d orbitals (figure 10). The oxygen molecule gives rise to a defect level 2.36 eV below the valence band.

The charged Zr vacancies do not result in the formation of an oxygen molecule. The 6 fold coordinated polyhedron below the vacancy has become 5 fold coordinated with the oxygen ion moving towards the vacancy site for the singly charged vacancy. The distance between the Ti and O ion has increased from 1.89 Å in the bulk to 2.65 Å in the defective structure. The doubly charged vacancy delocalises the two holes on four oxygen ions surrounding the vacancy. These four oxygen ions are all 2 coordinated in the defect structure, bridging a 6 coordinated Ti and a Zr ion. The triply charged defect delocalises the hole over two 2 coordinated oxygen ions, which bridge a Ti and Zr. The localised holes are on oxygen ions which are the same side of the vacancy and are separated by the Zr/Ca row. The quadruply charged vacancy has no holes associated with this vacancy and thus a delocalised state is prevented. The 4 fold coordinated polyhedron is displaced from the Ti row in these charged vacancies by 0.69 Å, 0.68 Å, 0.69 Å and 0.85 Å for the singly, doubly, triply and quadruply charged vacancy respectively. There are no defect levels present in the band gap for any of these charged vacancies.

*Table II: Summary of the effect of charge localisation on the defect structure of the Zr vacancy.*

| Charge state | Defective structure |
| --- | --- |
| $V^0_{Zr}$ | $O_2$ molecule<br>6 x 5 coordinated Ti<br>1 x 6 coordinated Ti |
| $V^-_{Zr}$ | 5 x 5 coordinated Ti<br>2 x 6 coordinated Ti |
| $V^{2-}_{Zr}$ | 5 x 5 coordinated Ti<br>1 x 6 coordinated Ti |
| $V^{3-}_{Zr}$ | 6 x 5 coordinated Ti |
| $V^{4-}_{Zr}$ | 5 x 5 coordinated Ti |

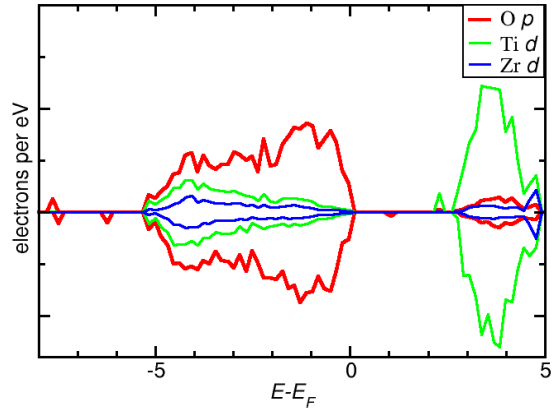

*Figure 10: (Color online) Partial density of states of the neutral Zr vacancy.*

There are three different chemical environments for the titanium ions and the relaxed vacancy structure is strongly dependent on the environment as table III shows. In the first environment the neutral and singly charged vacancies result in the formation of an oxygen molecule within the vacancy site, similar to that of the neutral zirconium vacancy, with bond lengths of 1.25 Å and 1.24 Å, respectively. The phonon frequency of these molecules are 37.05 THz and 37.59 THz (1235.90 cm$^{-1}$ and 1253.86 cm$^{-1}$) for the neutral and singly charged vacancies respectively. The isosurface of the neutral vacancy is shown in figure 8 at -0.01 eVÅ$^{-3}$ and clearly shows electron density in the middle of the two oxygen atoms. This build up of electron density indicates that a covalent bond has formed. The molecule formed by the singly charged vacancy has a similar electron density to that of the neutral vacancy $O_2$ molecule. The higher charged vacancies do not result in the formation of an oxygen molecule, however, the triply and quadruply charged vacancies lead to the formation of a 5 fold coordinated Ti polyhedron at the expense of the 4 fold coordinated polyhedron next to the vacancy.

The vacancy in the second Ti environment causes the formation of a 6 fold coordinated Ti polyhedron from the 4 fold coordinated polyhedron next to the vacancy site in all of the charge states. The quadruply charged vacancy gives rise to defect levels in the band gap formed from oxygen p orbitals at 0.15 eV and 0.28 eV above the valence band (figure 12). The 4 fold coordinated vacancy results in the formation of a 5 fold coordinated Ti polyhedron from the other 4 fold coordinated Ti polyhedron in the Ti row for all vacancy charge states. An additional 5 fold coordinated Ti polyhedron is formed in the Ti row below the vacancy in the neutral and doubly charged state. The presence of the doubly charged or lower charged vacancies forms an oxygen dumb-bell in the adjacent Ti row by two bridging oxygen ions at the expense of a 6 fold coordinated polyhedron. The presence of the 4 fold coordinated Ti vacancy results in the formation of three 5 fold coordinated Ti polyhedra when the vacancy is neutral or doubly charged.

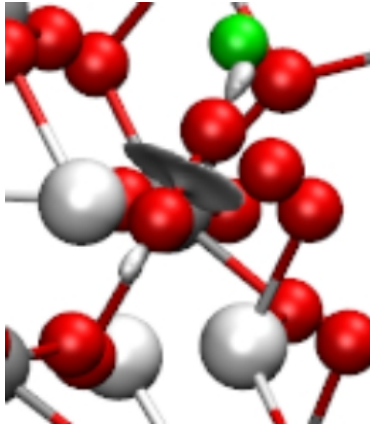

*Figure 11: (color online) Isosurface (-0.01 eVÅ⁻³) of the 6 coordinated Ti neutral vacancy showing electron density between the oxygen atoms viewed along the <001> direction*

*Table III: Summary of the effect of charge localisation and chemical environment, as described in text, on the defect structure of the Ti vacancies.*

| Charge state | 1$^{st}$ environment | 2$^{nd}$ environment | 3$^{rd}$ environment |
|---|---|---|---|
| $V^0_{Ti}$ | $O_2$ molecule | 6 coordinated Ti | 3 × 5 coordinated Ti |
| $V^-_{Ti}$ | $O_2$ molecule | 6 coordinated Ti | 2 × 5 coordinated Ti |
| $V^{2-}_{Ti}$ |  | 6 coordinated Ti | 3 × 5 coordinated Ti |
| $V^{3-}_{Ti}$ | 5 coordinated Ti | 6 coordinated Ti | 1 × 5 coordinated Ti |
| $V^{4-}_{Ti}$ | 5 coordinated Ti | 6 coordinated Ti | 1 × 5 coordinated Ti |

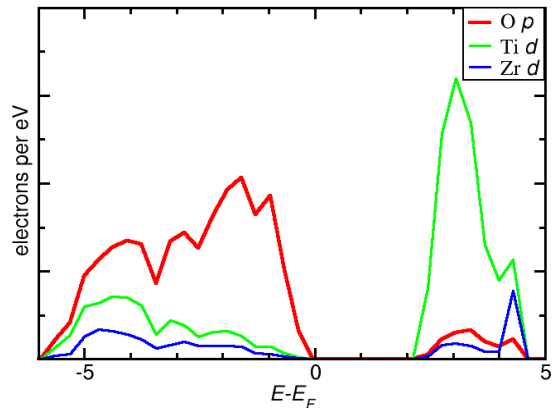

*Figure 12: (Color online) Partial density of states quadruply charged Ti vacancy in the 2$^{nd}$ environment.*

It is rare that DFT is able to describe hole localisation in oxide materials,[36] however it has been observed to correctly localise hole states at low coordinated sites on the MgO surface.[37] Analysis of the oxygen coordination in zirconolite shows that the oxygen ions are either 2 or 3 fold coordinated and thus providing sites for holes to localise. The calcium and first Ti vacancies localise holes on 2 fold coordinated oxygen ions, while the second Ti vacancy localises the holes on only 3 fold coordinated oxygen ions. The Zr and the 4 fold coordinated Ti vacancy localises the holes on both 2 and 3 fold coordinated oxygen ions.

### 3.2.2 Vacancy defect formation energies

The formation energy of the vacancy defects has been calculated to determine the stable charge state of the vacancy and the results are summarised in Table IV. The formation energy is calculated using equation 1 and is with respect to the chemical potential of the element and the electrons.

$$E_f(\mu_x, E_F) = (E_v - q\Delta V) - E_B + \mu_x + qE_{VBM} + q\Delta E_F \qquad (1)$$

Here $E_f$ is the formation energy, $E_v$ is the energy of the cell containing the vacancy, $q\Delta V$ is the potential alignment correction,[38] $E_B$ is the energy of the bulk cell, $\mu_X$ is the chemical potential of the isolated element in a vacuum ($\mu_X = E_X$, with the oxygen atom in a triplet state), $E_{VBM}$ is the energy of the valence band maximum and $E_F$ is the Fermi level which acts as an electron reservoir and is set to the valence band maximum in these calculations.

*Table IV: Summary of the formation energy for the vacancies in different charge states in bulk zirconolite. The formation energies are with respect to the chemical potential of the isolated atom in vacuum and the electron chemical potential set to the valence band maximum.*

| Charge state | O bridging two 6 coordinated Ti | O short Ti Zr bridge | O long Ti Zr bridge | 3 coordinated bridging O | Ca | Zr | 6 coordinated Ti | 6 coordinated Ti | 4 coordinated Ti |
|---|---|---|---|---|---|---|---|---|---|
| Neutral (eV) | 3.75 | 5.78 | 5.04 | 5.43 | 8.16 | 20.4 | 17.89 | 17.36 | 18.58 |
| Singly charged (eV) | 1.1 | 2.99 | 2.58 | 1.43 | **7.84** | 20.94 | 18.02 | 16.69 | 18.37 |
| Doubly charged (eV) | **-1.32** | **0.49** | **-0.2** | **-1.19** | 7.90 | 19.35 | 17.54 | **16.33** | 18.46 |
| Triply charged (eV) | | | | | | 19.17 | **17.32** | 16.60 | **16.35** |
| Quadruply charged (eV) | | | | | | **19.05** | 17.66 | 16.91 | 16.48 |

Table IV shows that for this position of the $E_F$ that the neutral defects are the least stable for all species, apart from the Zr vacancy which 0.5 eV more stable than the singly charged defect. The chemical environment of the 6 coordinated Ti has an effect on the stable vacancy charge state with the Ti in the 6 coordinated row favouring the triple charge state, while the Ti in the alternating row favours the doubly charge state. The value of $E_f$ for the oxygen vacancies are in good agreement with the formation of oxygen vacancies in $ZrO_2$.[39]

**3.3 Frenkel defects**

The energies of the isolated interstitials and vacancies were used to calculate the Frenkel pair energies, equation 2, for each species present in zirconolite. The Frenkel pair energies are summarised in table V. Only the neutral O pairs are considered here.

$$E_{FP} = (E_i - q\Delta V) + (E_v - q\Delta V) + (2E_B) \qquad (2)$$

where $E_{FP}$ is the Frenkel defect energy, $E_i$ is the energy of the cell containing the interstitial, $E_v$ is the energy of the cell containing the vacancy, $q\Delta V$ is the potential alignment correction[38] and $E_B$ is the energy of the bulk cell.

*Table V: Frenkel pair energies of the species within zirconolite. * denotes energy of Zr interstitial located in <010> channel is used to calculate the Frenkel pair energy. Refer to the text for more description of oxygen vacancy sites.*

| Frenkel Pair | Defect Energy (eV) |
|---|---|
| $O^0$ + two 6 coordinated Ti bridging $V_O^0$ | 1.15 |
| $O^0$ + Ti Zr short bridging $V_O^0$ | 3.18 |
| $O^0$ + Ti Zr long bridging $V_O^0$ | 2.44 |
| $O^0$ + 3 coordinated bridging $V_O^0$ | 2.83 |
| $Ca^{2+}$ + $V_{Ca}^{2-}$ | 1.97 |
| $Zr^{2+}$* + $V_{Zr}^{2-}$ | 8.22 |
| $Ti^{2+}$ + 1st 6 coordinated $V_{Ti}^{2-}$ | 5.11 |
| $Ti^{2+}$ + 2nd 6 coordinated $V_{Ti}^{2-}$ | 3.98 |
| $Ti^{2+}$ + 4 coordinated $V_{Ti}^{2-}$ | 6.03 |

Table V shows that the Frenkel defect energies are relatively low and that they are consistent with other complex materials, with the oxygen Frenkel pair energies ranging from 2.8-5.3 eV in pyrochlores[40] and a Zr Frenkel pair energy of 11.16 eV in zircon.[41]

## 4. Conclusions

We have used DFT to study intrinsic defects in zirconolite, a complex ceramic that is a promising candidate for encapsulating high level radioactive waste. The complex crystal structure and the range of possible defect oxidation states means that this material is challenging to study using both experimental and modelling techniques.

We have used an efficient random search technique to identify the lowest energy conformation for interstitials in the zirconolite unit cell. This method enabled the identification of stable defect structures that would have been difficult to predict by traditional methods. The neutral O interstitial resulted in the removal of the 4 fold coordinated Ti polyhedra, whereas the O$^-$ interstitial increased the coordination of a Ti ion from 4 to 5. The Ca and Ti interstitials favour sites in the <010> channels between the ions, whereas the Zr interstitial displaced a Ti ion from its lattice site to form a substitutional Zr ion and a Ti interstitial.

Vacancies in zirconolite were also shown to have interesting structures, which were sometimes strongly dependent on the oxidation state. In particular, the neutral Zr vacancy resulted in the formation of an $O_2$ molecule, whereas the singly charged vacancy did not result in molecule formation, but it did cause significant displacement of one oxygen atom on a neighbouring 6 fold coordinated Ti ion. Oxygen molecules were also observed to form near some Ti vacancies, those associated with the 6 fold coordinated Ti ion, but not in others. The chemical environment of the O vacancies was found to have a strong effect on the formation energies of oxygen Frenkel pairs, which range between 1.15 eV and 3.18 eV for different chemical environments. The low Frenkel pair formation energies could explain why zirconolite requires a small number of displacements per atom (dpa) to become amorphous[2] as the defects are relatively stable.

## 5. Acknowledgements

We are grateful to Chris Pickard, Alex Shluger, Keith McKenna and Mathew Watkins for useful discussion. The authors acknowledge the use of the UCL *Legion* High Performance Computing Facility, and associated support services, in the completion of this work. We acknowledge support from EPSRC via the DIAMOND consortium (EP/F055412/1).

## 6.References